\documentclass[11pt,a4paper]{article}

\usepackage{float}
\usepackage{graphicx}
\usepackage{amsmath}
\usepackage{amsfonts}
\usepackage{latexsym}
\usepackage[dvips,a4paper]{hyperref}
\newcommand{\mf}[1]{\mathfrak{#1}}

\newcommand{\beq}{\begin{equation}}
\newcommand{\eeq}{\end{equation}}
\newcommand{\proof}{\noindent {\sl Proof.\ }}

\newcommand{\RR}{\mathbb{R}}

\begin{document}

\begin{center}
{\Large \bf
On Lagrangian and Hamiltonian systems with \\[2mm]
homogeneous trajectories}
\vspace{1.5cm}

{\large
G\'abor  Zsolt T\'oth}
\vspace{1cm}

{\it
KFKI Research Institute for Particle and Nuclear Physics,\\
Hungarian Academy of Sciences, P.O.B.\ 49, 1525 Budapest,
Hungary}\\[8mm]
{\tt email:\ \
tgzs@rmki.kfki.hu, tgzs@cs.elte.hu}
\end{center}
\vspace{1.5cm} 

\begin{center}
{\bf Abstract}
\end{center}
\vspace{0.5cm}

\noindent
{\small 
Motivated by various results on homogeneous geodesics of Riemannian spaces,
we study homogeneous trajectories, i.e.\ trajectories which are orbits of a one-parameter 
symmetry group, of Lagrangian and Hamiltonian systems.
We present 
criteria under which an orbit of a one-parameter subgroup of a symmetry group $G$ is a 
solution of the Euler-Lagrange or Hamiltonian equations. In particular, we generalize the `geodesic lemma' known in Riemannian geometry to Lagrangian and Hamiltonian systems. 
We present results on the existence of homogeneous trajectories of Lagrangian systems.
We study Hamiltonian and Lagrangian g.o.\ spaces, i.e.\ homogeneous spaces $G/H$ with $G$-invariant Lagrangian or Hamiltonian functions on which every solution of the equations of motion is homogeneous.
We show  that the Hamiltonian g.o.\ spaces are related to the functions that are invariant under the coadjoint action of $G$. Riemannian g.o.\ spaces thus 
correspond to special $Ad^*(G)$-invariant functions. 
An $Ad^*(G)$-invariant function that is related to a g.o.\ space also serves as a potential for the mapping called `geodesic graph'.
As illustration we discuss the Riemannian g.o.\ metrics on $SU(3)/SU(2)$.}
\vspace{1.5cm}



\thispagestyle{empty}

\newpage

\section{Introduction}

Let $M$ be a Riemannian manifold. 
A geodesic in $M$ is called {\it homogeneous}  if it is the orbit of a one-parameter 
group of isometries of  $M$.
A homogeneous Riemannian manifold 
$M=G/K$, where $G$ is a connected Lie group 
and $K$ is a closed subgroup,
is a {\it geodesic orbit (g.o.) space} with respect to $G$, 
if every geodesic in it is the orbit of a one-parameter 
subgroup of $G$. 

The homogeneous space $M=G/K$
is called a {\it reductive space}, if there exists a direct sum decomposition (called {\it reductive decomposition}) $\mf{g}=\mf{m}\oplus\mf{k}$ 
of the Lie algebra of $G$,
where $\mf{m}$ is an 
$ad(K)$-invariant linear subspace of $\mathfrak{g}$ and $\mf{k}$ is the Lie algebra of $K$. It is known that 
all Riemannian homogeneous spaces are reductive. If $M=G/K$ is Riemannian and there exists 
a reductive decomposition $\mf{g}=\mf{m}\oplus\mf{k}$  such that
each geodesic in $M$  starting at the origin $o\in M$ 
is an orbit of a one-parameter subgroup of $G$ generated by some element of $\mf{m}$,
then $M$  is called a 
{\it naturally reductive space} with respect to $G$, and $\mf{m}$ is called a {\it natural complement}. The origin $o$ is the image of $K$ by the canonical projection $G\to G/K$.

Obviously, every naturally 
reductive space is a g.o.\ space as well. It was believed some decades ago
that the converse is also true, i.e.\ every  
g.o.\ space is isometric to some naturally reductive space. A counter example,
however, was found by A.\ Kaplan \cite{Kaplan},
initiating the extensive
study of g.o.\ spaces  
\cite{R}-\cite{AN}.
Pseudo-Riemannian g.o.\ spaces were also investigated recently \cite{DK1,DK2,DK3}. 
Before Kaplan's example appeared, J.\ Szenthe discovered a geometrical background for the situation when a g.o.\ space is not naturally reductive \cite{Szenthe1}, not knowing whether such a situation can be realized or not. This result
had considerable influence on the later studies.

In general, it is possible that a homogeneous Riemannian space $M=G/K$ is not naturally reductive with respect to $G$, but one can take other groups  
$G'$ and $K'$ so that $M=G'/K'$ and $M$ is naturally reductive with respect to $G'$. The same situation can occur for g.o.\ spaces as well.
It is also possible in some cases that a g.o.\ space can be made naturally reductive by taking a different symmetry group $G'$, but there also exist 
g.o.\ spaces for which this is not possible, i.e.\ which are {\it in no way} naturally reductive. Kaplan's example is of the latter type.

Since Riemannian (and pseudo-Riemannian) manifolds can be viewed as a special class of the manifolds with a Lagrangian or Hamiltonian function, it is interesting to consider the generalization of the g.o.\ property to
homogeneous spaces with invariant Lagrangian and Hamiltonian functions and to ask whether the known results for the Riemannian spaces can be generalized, and whether 
the techniques of Lagrangian or Hamiltonian dynamics can be used for the study of Riemannian g.o.\ spaces. In this paper we present the results that we obtained in relation to these questions. 

A subject closely related to the study of g.o.\ spaces is the characterization of the homogeneous 
geodesics in Riemannian manifolds. Homogeneous geodesics are of interest also in Finsler geometry, pseudo-Riemannian geometry and in dynamics.  We refer the reader to \cite{Hermann}-\cite{CM}
and further references therein.
The present paper is also concerned with the characterization of homogeneous trajectories in  Lagrangian and Hamiltonian dynamical systems, partly because this is necessary for the study of dynamical systems that have the g.o.\ property.
In the physics literature the homogeneous geodesics are usually called relative equilibria, therefore we shall also use this term, along with the term homogeneous trajectory. 
We mention that another name for homogeneous geodesics that appears in the literature is stationary geodesic.
At times we shall use the terms Lagrangian space and Hamiltonian space for Lagrangian and Hamiltonian dynamical systems, in analogy with the term Riemannian space.

The paper is organized as follows.
In section \ref{sec.2} we discuss the case of Lagrangian systems. We describe criteria for an orbit of a one-parameter subgroup  to be a solution of the Euler-Lagrange equations, including the Lagrangian version of the `geodesic lemma'.
We also present results concerning the existence of relative equilibria.

In section \ref{sec.3} we discuss the case of Hamiltonian systems. We describe criteria for an orbit of a one-parameter subgroup  to be a solution of the Hamiltonian equations, including the Hamiltonian version of the geodesic lemma. Then we turn to the characterization of Hamiltonian g.o.\ spaces. In particular, we show  that the Hamiltonian g.o.\ spaces are closely related to the functions which are invariant under the coadjoint action of $G$.
Riemannian g.o.\ spaces correspond, of course, to special $Ad^*(G)$-invariant functions. 
Naturally reductive metrics, in particular, are known to correspond to quadratic $Ad^*(G)$-invariant polynomials \cite{Kostant,DAtriZiller}.
An $Ad^*(G)$-invariant function that is related to a g.o.\ space also serves as a potential for the mapping called {\it geodesic graph},
which was introduced originally by Szenthe \cite{Szenthe1} and which has proved to be useful  
for the description of Riemannian g.o.\ spaces. We present certain results on geodesic graphs, and then we
describe a criterion based on the relation between g.o.\ spaces and $Ad^*(G)$-invariant functions that can be used to find g.o.\ Hamiltonians or metrics.
We also describe a generalization of the notion of Hamiltonian g.o.\ space.

In section \ref{sec.4} we discuss the two-parameter family of Riemannian g.o.\ metrics on\\ $SU(3)/SU(2)$ for the illustration of the results of section \ref{sec.3}. We calculate the geodesic graph in a new way, utilizing the relation between g.o.\ spaces and $Ad^*(G)$-invariant functions.

\section{Lagrangian systems with homogeneous trajectories}
\label{sec.2}

Let $M$ be a connected manifold with a Lagrangian function $L:TM\to\mathbb{R}$ on it. 
The {\it Euler-Lagrange equation} for a curve $\gamma:\ I\to M$, where $I$ is an interval, is
\beq
\frac{\partial L}{\partial x^i}(\gamma(t),\dot{\gamma}(t))=\frac{d}{dt}\left(\frac{\partial L}{\partial v^i}(\gamma,
\dot{\gamma})\right)(t) \quad\quad \forall t\in I,
\eeq
or, expanding the right hand side,
\beq
\label{EL}
\frac{\partial L}{\partial x^i}(\gamma(t),\dot{\gamma}(t))=\frac{\partial^2L}{\partial x^i\partial v^j}(\gamma(t),\dot{\gamma}(t))\dot{\gamma}^j(t)+
\frac{\partial^2L}{\partial v^i\partial v^j}(\gamma(t),\dot{\gamma}(t))\ddot{\gamma}^j(t).
\eeq
Here and throughout the paper we use the Einstein summation convention for indices of coordinates related to $M$. 
In the special case when $L$ is the quadratic form corresponding to a Riemannian or pseudo-Riemannian metric,
a solution $\gamma:\ I\to M$ of the Euler-Lagrange equations is a geodesic with affine parametrization.

The Lagrangian is {\it regular} if the bilinear form $\frac{\partial^2 L}{\partial v^i \partial v^j}(x,v)$ is nondegenerate for any $(x,v)\in TM$. 
The regularity of a Lagrangian implies that
the solution of the Euler-Lagrange equations is unique for given initial data $(x,v)\in TM$.
If a Lagrangian corresponds to a metric, then it is regular.

In the following we assume that $L$ is invariant under
the action of a connected Lie group $G$ on $TM$ induced by an action of $G$ on $M$. 
We denote the Lie derivative with respect to a vector field $Z$ as $\mathcal{L}_Z$. 
We use the notation $\circ$ for the composition of two functions, i.e.\ if $f$ and $g$ are two functions, then $f\circ g$ is the function for which $(f\circ g)(x)=f(g(x))$.
 
In the derivation of the results of this section  the Euler-Lagrange equation,
an equation expressing the invariance of $L$ and equations characterizing the 
velocity and acceleration of orbits have important role.\\

Let $Z_a : M\to TM$  and $\hat{Z}_a : TM\to TTM$, where $a\in\mathfrak{g}$, be the infinitesimal generator
vector fields for the action of $G$ on $M$ and $TM$, respectively. Their coordinate form 
is
\beq
Z_a(x)=\frac{\partial\phi_a^i}{\partial\tau}(0,x)\frac{\partial}{\partial x^i},\qquad  x\in M
\eeq
and
\beq
\hat{Z}_a(x,v)=\frac{\partial\phi_a^i}{\partial\tau}(0,x)\frac{\partial}{\partial x^i}+
\frac{\partial^2\phi_a^i}{\partial\tau\partial x^j}(0,x)v^j\frac{\partial}{\partial v^i},\qquad (x,v)\in TM,
\eeq
where  $\phi_a : \mathbb{R}\times M\to M$ is the action of the one-parameter subgroup generated by $a\in\mathfrak{g}$ and $\tau$ denotes the first variable of $\phi_a$.

The invariance of $L$ under the action of $G$ implies the following 
{\it symmetry condition}: 
\beq
\label{sym}
\mathcal{L}_{\hat{Z}_a}L(x,v)=\frac{\partial L}{\partial x^i}(x,v)\frac{\partial\phi_a^i}{\partial\tau}(0,x)+\frac{\partial L}{\partial v^i}(x,v)\frac{\partial^2\phi_a^i}{\partial\tau\partial x^j}(0,x)v^j=0,
\eeq
where $a\in\mathfrak{g}$. This equation holds for all $(x,v)\in TM$.\\

The orbit of the one-parameter subgroup generated by $a\in\mathfrak{g}$ in $M$ with initial point $x$ is 
the curve $\gamma : I\to M,\ t\mapsto \phi_a(t,x)$. 
For the velocity
\beq
\dot{\gamma}(t)=\frac{\partial \phi_a}{\partial \tau}(t,x)
\eeq
of this orbit
the equation
\beq
\label{v1}
\dot{\gamma}^i(t)=\frac{\partial\phi_a^i}{\partial x^j}(t,x)\dot{\gamma}^j(0)=
\frac{\partial\phi_a^i}{\partial x^j}(t,x)\frac{\partial\phi_a^j}{\partial\tau}(0,x)
\eeq
holds because of the group property.
For the acceleration we have
\beq
\label{v2}
\ddot{\gamma}^i(t)=\frac{\partial^2\phi_a^i}{\partial\tau\partial x^j}(t,x)\frac{\partial\phi_a^j}{\partial\tau}(0,x)=
\frac{\partial^2\phi_a^i}{\partial\tau^2}(t,x).
\eeq
\vspace{0.5cm}

\noindent
{\bf Theorem 2.1}\ {\it  
The orbit of a one-parameter subgroup of $G$ starting at $x\in M$ is a solution of the Euler-Lagrange equations corresponding to the (not necessarily regular) Lagrangian $L$
if and only if $x$ is a critical point of the function $L\circ Z_a$, i.e.\
\beq
\label{1}
d (L\circ Z_a)(x)=0,
\eeq
where $Z_a$ is the generator vector field of the subgroup.
}\\

\proof Because of  the invariance of the Lagrangian  an orbit of a one-parameter symmetry group is a solution of the Euler-Lagrange equations if and only if it satisfies
the Euler-Lagrange equations at the initial point.
First, let us assume that the orbit is a solution of the Euler-Lagrange equations. 
Differentiating the symmetry condition (\ref{sym}) with respect to $v^j$ yields
\begin{eqnarray}
0 & = & \frac{\partial}{\partial v^j}\mathcal{L}_{\hat{Z}_a}L(x,v)=\frac{\partial^2L}{\partial x^i\partial v^j}(x,v)\frac{\partial\phi_a^i}{\partial\tau}(0,x)\nonumber \\
&& + \frac{\partial^2L}{\partial v^i\partial v^j}(x,v)\frac{\partial^2\phi_a^i}{\partial\tau\partial x^k}(0,x)v^k+
\frac{\partial L}{\partial v^i}(x,v)\frac{\partial^2\phi_a^i}{\partial\tau\partial x^j}(0,x).
\label{sym1}
\end{eqnarray}
Substituting  the right hand side of (\ref{v2}) for $\ddot{\gamma}$ in the Euler-Lagrange equation (\ref{EL}) at $t=0$ gives
\beq
\label{EL1}
\frac{\partial L}{\partial x^j}(x,v)=\frac{\partial^2L}{\partial x^i\partial v^j}(x,v)\frac{\partial\phi_a^i}{\partial\tau}(0,x)+
\frac{\partial^2L}{\partial v^i\partial v^j}(x,v)\frac{\partial^2\phi_a^i}{\partial\tau\partial x^j}(0,x)\frac{\partial\phi_a^j}{\partial\tau}(0,x),
\eeq
where $v=\dot{\gamma}(0)$.
Setting $v=\dot{\gamma}(0)$ also in (\ref{sym1}) and subtracting from (\ref{EL1}) gives
\beq
\label{krit}
\frac{\partial L}{\partial x^j}(x,v)+\frac{\partial L}{\partial v^i}(x,v)\frac{\partial^2\phi_a^i}{\partial\tau\partial x^j}(0,x)
=0,
\eeq
where $v=\dot{\gamma}(0)$,
which is just the coordinate form of (\ref{1}).
Considering the reverse direction of the statement, it is clear now that if (\ref{krit}) and (\ref{sym1}) hold, then (\ref{EL1}) follows. 
$\Box$\\

A similar theorem is stated in \cite{Lewis} (see also \cite{CM}). However, our proof is different from those given in \cite{Lewis} and \cite{CM}.
The function $L\circ Z_a$ is called {\it augmented Lagrangian} in \cite{Lewis} and {\it locked Lagrangian} in \cite{CM}.
\\

\noindent
{\bf Definition 2.2}\ 
An element  $a$ of $\mf{g}$ is called a {\it relative equilibrium vector} at $x\in M$ if the
orbit of the  one-parameter subgroup of $G$ generated by $a$ and starting at $x$  is a solution of the Euler-Lagrange equations.\\

In Riemannian geometry the interesting relative equilibrium vectors are, of course, those which generate orbits that are not single points in $M$. We note that in Riemannian geometry the relative equilibrium vectors are usually called {\it geodesic vectors}.

The set of relative equilibrium vectors at $x$ is invariant
under $G_x$, the stabilizer of $x$. If $gx=y$ for some $x,y\in M$ and 
$g\in G$, then the set of relative equilibrium vectors at $y$ can be obtained from that at $x$ by 
the adjoint action of $g$.

As regards the existence of relative equilibria, the following corollary of 
theorem 2.1 can be 
stated.\\

\noindent
{\bf Theorem 2.3}\ {\it
Let $M$, $G$, $L$ be as in the theorem 2.1 and let $M$ be compact. 
For any  $a\in\mf{g}$ there exists at least one solution of the Euler-Lagrange equations
which is
the orbit of the one-parameter subgroup generated by $a$.
If there
exists an $a\in\mf{g}$ such that
$Z_{a}(x)\ne 0\ \  \forall x\in M$, then  there exists at least one solution of the Euler-Lagrange equations 
which is
the orbit of the one-parameter subgroup generated by $a$ and is not a single point in $M$. If, in addition,  
$M$ is also homogeneous with respect to the action of $G$, then there exists  at least 
one nonzero relative equilibrium vector at every point in $M$, which generates an orbit that is not a single point.}\\

This result can be found e.g.\ in \cite{Lacomba} (proposition 5.2) for the special case of Lagrangians that describe geodesic motion in 
Riemannian manifolds.

In the rest of this section we consider the case when $M$ is a homogeneous space. 
For a homogeneous space $M=G/K$ there is a linear map
$f_x:\ \mathfrak{g}\to T_xM,\ a\mapsto Z_a(x)$ for each point $x\in M$. 
We use the notation $f$ for $f_o$ (i.e.\ we omit the subscript $o$ denoting the origin in $G/K$).  

The dual of a vector space $V$ will be denoted by $V^*$. The contraction (or natural pairing) between 
$V$ and $V^*$ will be denoted in the following way: $(w|v)$, where $w\in V^*$ and $v\in V$. The transpose
of a linear map $A:V\to W$ will be denoted by $A^*$ (it is defined as  $A^*: W^*\to V^*,
\ w\mapsto w\circ A$).

The following lemma, which concerns homogeneous manifolds with invariant Lagrang-ians and is the generalization of the known `geodesic lemma' for the Riemannian case
\cite{K-V} (see also for example \cite{KSZ,CSG1,KowNikc}), gives a condition for an element of $\mf{g}$ to be a relative equilibrium vector at $o$. This is a local condition in the sense that it is given in terms of $L$ restricted to $T_oM$, the elements of $\mf{g}$, and the values of the infinitesimal 
generator vector fields at $o$. In Riemannian geometry the geodesic lemma has proved to be very useful in the study of homogeneous geodesics.\\

\noindent
{\bf Lemma 2.4 (Geodesic lemma)}\ {\it 
Let $M=G/K$ be a homogeneous space with a $G$-invariant Lagrangian 
$L:TM\to\mathbb{R}$. An element $a \in \mathfrak{g}$ is a relative equilibrium vector at $o$ if and only if 
\beq
\label{c2}
(\ dL_o(f(a))\ |\ f([a,b])\ )=0\qquad \forall b\in \mf{g},
\eeq
where $L_o$ is $L$ restricted to $T_oM$.
In particular, if $L$ corresponds to a Riemannian metric, then (\ref{c2}) takes the form 
\beq
\langle f([a,b])\ ,\ f(a)\rangle =0\qquad \forall b\in \mf{g},
\eeq
or, equivalently,
\beq
\label{r}
\langle [a,b]_{\mf{m}}\ ,\ a_{\mf{m}}\rangle=0\qquad \forall b\in \mf{g},
\eeq
where the index $\mf{m}$ denotes the $\mf{m}$-component related to a reductive decomposition $\mf{g}=\mf{k}\oplus\mf{m}$, and $\mf{m}$ is assumed to be identified with $T_oM$ by $f$.}\\

\proof
Let us assume first, that $a$ is a relative equilibrium vector. 
(\ref{1}) in theorem 2.1 is equivalent to $\mathcal{L}_{Z_b}(L\circ Z_a)(o)=0\ \ \forall b\in\mathfrak{g}$.
In coordinate form
\begin{eqnarray}
\mathcal{L}_{Z_b}(L\circ Z_a)(o)=\frac{\partial \phi_b^i}{\partial \tau}(0,o)\frac{\partial L}{\partial x^i}(o,\frac{\partial\phi_a}{\partial\tau}(0,o))
\nonumber \\
\label{krit3}
+\frac{\partial \phi_b^i}{\partial\tau}(0,o)\frac{\partial L}{\partial v^j}(o,\frac{\partial\phi_a}{\partial\tau}(0,o))
\frac{\partial^2\phi_a^j}{\partial\tau\partial x^i}(0,o)=0.
\end{eqnarray}
Taking the symmetry condition (\ref{sym}) at the point
$(o,\frac{\partial\phi_a}{\partial\tau}(0,o))$
we get
\begin{eqnarray}
\mathcal{L}_{\hat{Z}_b}L(o,\frac{\partial\phi_a}{\partial\tau}(0,o))=
\frac{\partial \phi_b^i}{\partial \tau}(0,o)\frac{\partial L}{\partial x^i}(o,\frac{\partial\phi_a}{\partial\tau}(0,o)) \nonumber\\ 
+\frac{\partial L}{\partial v^i}(o,\frac{\partial\phi_a}{\partial\tau}(0,o))\frac{\partial^2\phi_b^i}{\partial\tau
\partial x^j}(0,o)\frac{\partial \phi_a^j}{\partial \tau}(0,o)=0.
\label{symx}
\end{eqnarray}
Subtracting these two equations gives
\beq
\label{krit2}
\frac{\partial L}{\partial v^j}(o,\frac{\partial\phi_a}{\partial\tau}(0,o))
\left[\frac{\partial \phi_b^i}{\partial \tau}(0,o)\frac{\partial^2\phi_a^j}{\partial\tau\partial x^i}(0,o)-
\frac{\partial \phi_a^i}{\partial \tau}(0,o)\frac{\partial^2\phi_b^j}{\partial\tau\partial x^i}(0,o)\right]=0,
\eeq
which is the coordinate expression for (\ref{c2}). 
Conversely, assuming that (\ref{krit2}) holds and using (\ref{symx}) one obtains (\ref{krit3}).
The second part of the lemma concerning the Riemannian case follows obviously from the first part. $\Box$\\

The formula (\ref{r}) for Riemannian spaces is well known and is also a generalization of Arnold's result about homogeneous geodesics of left-invariant metrics on Lie groups \cite{Arnold}.

Let $r:\mathbb{R} \to \mathfrak{g}$ be the adjoint orbit starting at $a$ and generated by $b$. 
$f([a,b])$ is the tangent vector of the curve $f\circ r$ at the point $f(a)$. Equation (\ref{c2}) means that the derivative of $L_o$ at $f(a)$ along this tangent vector is $0$.\\

The following theorems 2.5 and 2.6 are about the existence of relative equilibria.\\

\noindent
{\bf Theorem 2.5}\ {\it
Let $M=G/K$ be a homogeneous space with a $G$-invariant Lagrangian 
$L:TM\to\mathbb{R}$.
If $G$ is compact, then each adjoint orbit of $G$ contains at least one relative equilibrium vector at $o$, and
each adjoint orbit of $G$
that is not contained entirely  by  $\mf{k}$ 
contains at least one relative equilibrium vector at $o$ which generates an orbit that is not a single point. }\\

\proof Any adjoint orbit $O$ of $G$ is compact.  $f(O)$ is also compact and $L_o$ is 
continuous on it, thus there exists at least one $\tilde{v}\in f(O)$  so 
that
$L_o|_{f(O)}$ is minimal or maximal at $\tilde{v}$. Because of this extremality the derivative of $L_o$ is zero at $\tilde{v}$ along any curve that lies in $f(O)$ and passes through $\tilde{v}$.  
It is clear from the remark after the proof of the geodesic lemma  
that any element of $f^{-1}(\tilde{v})\cap O$ is a relative equilibrium vector at $o$.

If an adjoint orbit $O$  is not contained entirely by $\mf{k}$, then  $f(O)\ne \{ 0 \}$, thus there exists at 
least one $\tilde{v}\in f(O)$  so 
that $\tilde{v}\ne 0$ and
$L_o|_{f(O)}$ is minimal or maximal at $\tilde{v}$. 
Any element of $f^{-1}(\tilde{v})\cap O$ is a relative equilibrium vector at $o$ that generates an orbit that is not a single point.
$\Box$\\

\noindent
{\bf Theorem 2.6}\ {\it
Let $M=G/K$ be a homogeneous space with a $G$-invariant Lagrangian 
$L:TM\to\mathbb{R}$.
If $G$ is solvable and the image space of $dL_o|_{T_oM\setminus \{0\} }$ contains vectors of
arbitrary direction, than there exists at least one relative equilibrium vector at $o$, which generates an orbit that is not a single point.}\\

\proof Consider the derived series of $\mf{g}$, i.e.\ the sequence
$$\mf{g}^{(0)}\supset\mf{g}^{(1)}\supset \dots \supset{g}^{(i)}\supset \dots,$$
where $\mf{g}^{(0)}=\mf{g}$ and $\mf{g}^{(i)}=[\mf{g}^{(i-1)},\mf{g}^{(i-1)}]$
for $i=1,2,\dots$.
Because of the solvability of $G$, the derived series strictly decreases and 
ends in the null space. Consequently, there exists an index $r\ge 0$ such that
$f(\mf{g}^{(r)})=T_oM$, but $f(\mf{g}^{(r+1)})$ is a proper subspace of 
$T_oM$. The connected subgroup $G^{(r)}$ corresponding to $\mf{g}^{(r)}$ still acts transitively on $M$, therefore
it is necessary and sufficient for a vector to be a relative equilibrium vector that (\ref{c2}) hold for all $b \in \mf{g}^{(r)}$. 
The condition imposed on $dL_o$ in the theorem ensures that there exists an $\tilde{v}\in T_oM\setminus
\{ 0\}$ such that $(\ dL_0(\tilde{v})\ |\ f([\mf{g}^{(r)},\mf{g}^{(r)}])\ )=0$, implying that
any element of $f^{-1}(\tilde{v})\cap\mf{g}^{(r)}$ is a relative equilibrium vector.
$\Box$\\

This theorem is similar to some parts of proposition 3 of \cite{KSZ}. It is clear from the proof that the solvability of $G$ is not necessary, it can be replaced by the weaker condition that there exists an element $\mf{g}^{(r+1)}$ of the derived series of $\mathfrak{g}$ such that 
$f(\mf{g}^{(r+1)})$ is a proper subspace of 
$T_oM$.

The condition of regularity has not been imposed  on the Lagrangians so far.
It is assumed, however, in the following two  propositions 2.8 and 2.10, which characterize 
Lagrangian g.o.\ spaces.\\

\noindent
{\bf Definition 2.7}\ 
Let $M=G/K$ be a homogeneous space and let
$L:M\to\mathbb{R}$ be a $G$-invariant Lagrangian function. $(M,L)$ is a called a {\it Lagrangian geodesic 
orbit (g.o.) space with respect to $G$}, if every solution of the Euler-Lagrange equations corresponding to $L$ is an orbit of a one-parameter subgroup of $G$.\\

In other words, a Lagrangian g.o.\ space is defined by the property that every solution of the Euler-Lagrange equations is a relative equilibrium.  
In the general Lagrangian mechanical context one could introduce a new name instead of `geodesic orbit space', since the latter bears a reference to 
Riemannian geometry. In the present paper, however, we shall not introduce such a name. This applies also to the 'geodesic lemma' and to the `geodesic graph' defined below.\\
\newpage

\noindent
{\bf Proposition 2.8}\
{\it  Let $M=G/K$ and $L$ be as in definition 2.7, and assume that $L$ is regular.
The Lagrangian dynamical system $(M,L)$ has the g.o.\ property with respect to $G$ if and only if for all
$v\in T_oM$ there exists an $a\in\mathfrak{g}$ such that $f(a)=v$ and $a$ is a relative equilibrium vector.}\\

\noindent
{\bf Definition 2.9}\
Let $(M=G/K,L)$ be a Lagrangian system that has the g.o.\ property with respect to $G$. A  
mapping $\xi: T_oM \to \mathfrak{g}$ 
with the properties that $f(\xi(v))=v$ and  $\xi(v)$ is a relative equilibrium vector at $o$ for all $v\in T_oM$
is called a 
{\it geodesic graph}. Obviously, there exists at least one  geodesic graph for every Lagrangian system that has the  g.o.\ property.
$f(\xi(v))=v$ means that the velocity of the orbit generated by $\xi(v)$ is $v$ at $o$.
\\

In Riemannian geometry the geodesic graph is very useful for studying g.o.\ spaces.
Important results about its properties were obtained in \cite{Szenthe1,KowNikc}.

It follows directly from the definition of naturally reductive metrics in the 
Introduction and from 2.9 
that the naturally reductive spaces are precisely those Riemannian g.o.\ spaces that 
admit a $K$-equivariant linear geodesic graph.
In order to see this in detail, assume first that
$M=G/K$ is a naturally reductive space with the natural reductive decomposition $\mf{g}=\mf{k}\oplus\mf{m}$. Then $f|_\mf{m}$ is a linear bijection between $\mf{m}$ and $T_oM$, and its inverse $\xi=(f|_{\mf{m}})^{-1}$ obviously has the property $f(\xi(v))=v$. 
$\xi$ is also $K$-equivariant, 
since $\mf{m}$ is an $Ad(K)$-invariant subspace of $\mf{g}$.
The natural reductivity of $M$ implies that for any $v\in T_oM$ there is an $a\in \mf{m}$ so that the orbit generated by $a$ and starting at $o$ coincides with the geodesic with initial velocity $v$. However, the initial velocity of the orbit generated by $a$ is $f(a)$, therefore $a=\xi(v)$. This shows that $\xi$ is a $K$-equivariant linear geodesic graph.  
Conversely, if $M$ is a Riemannian g.o.\ space and $\xi$ is a $K$-equivariant 
linear geodesic graph, then $\xi(T_oM)$ is an 
$Ad(K)$-invariant linear subspace of $\mf{g}$, due to the linearity and $K$-equivariance of $\xi$.
$\xi(T_oM)$ is complementary to $\mf{k}$, because $f(\mf{k})=0$ and $f(\xi(T_oM))=T_oM$. 
$\mf{g}=\mf{k}\oplus \xi(T_oM)$ is thus a reductive decomposition. By definition 2.9, 
for an arbitrary geodesic $\gamma$ starting at $o$ the Lie algebra element $\xi(v)\in \xi(T_oM)$,
where $v$ is the initial velocity of $\gamma$,
generates an orbit $\tilde{\gamma}$ that is also a geodesic with initial velocity $v$. 
Since geodesics are uniquely determined by their initial data, $\tilde{\gamma}$ and $\gamma$ coincide. This shows that the reductive decomposition $\mf{g}=\mf{k}\oplus \xi(T_oM)$ is also natural.

We note that there is a minor difference between our definition of the geodesic graph and the usual definition; in the usual definition one has a direct sum decomposition $\mathfrak{g}=\mathfrak{m}\oplus \mathfrak{k}$, and one
takes the $\mathfrak{k}$-component of $\xi(v)$ as the value of the geodesic graph at $v$, since the $\mf{m}$-component is uniquely determined by the property $f(\xi(v))=v$. In fact, in the literature $\mf{m}$ is often identified with $T_oM$ by $f$.
It is also usual in the literature to include in the definition of the geodesic graph the requirement that it should 
be $K$-equivariant. \\

The following consequence of proposition 2.8 and of the geodesic lemma, in particular of (\ref{c2}), applying to the special case $M=G$, 
is  well known \cite{Abr-Mars}.\\

\noindent
{\bf Theorem 2.10}\ 
{\it If $M=G$, i.e.\ $L$ is a regular left-invariant Lagrangian on $G$, 
then $(M,L)$ is a g.o.\ space with respect to $G$ if and only if $L_e=L|_{T_eG}$ (where $e$ is the unit element of $G$) is 
invariant under the adjoint action of $G$. Any function on $T_e G$ can be extended uniquely to a left-invariant 
function on $G$, therefore the Lagrangians on $G$ that have the g.o.\ property with respect to $G$ are in one-to-one correspondence with the regular $Ad$-invariant functions on 
$\mathfrak{g}$.}\\

\noindent
We note that
in the case $M=G$ the equation (\ref{c2}) expresses the $Ad(G)$-invariance of $L_e$.

In the next section we turn to the Hamiltonian formalism, which is better suited to the characterization of g.o.\ spaces than the Lagrangian formalism.

\section{Hamiltonian systems with homogeneous trajectories}
\label{sec.3}

Let $M$ be a manifold with a Hamiltonian function $H: T^*M\to\mathbb{R}$. 
We denote the Hamiltonian vector field generated by $H$ on the symplectic manifold $T^*M$ by $X_H$.
In coordinates $X_H$ is given by $X_H(x,p)=\left(\frac{\partial H}{\partial p_i}(x,p),-\frac{\partial H}{\partial x^i}(x,p)\right)$.
The {\it Hamiltonian equations} for a curve $\gamma:I\to T^*M$ are the following:
\beq
\label{HAM0}
X_H(\gamma(t))=\dot{\gamma}(t)\qquad \forall t\in I,
\eeq
or equivalently
\begin{eqnarray}
\label{HAM1}
\frac{\partial H}{\partial p_i}(x,p) & = & \dot{x}^i\\
\label{HAM2}
-\frac{\partial H}{\partial x^i}(x,p) & = & \dot{p}_i.
\end{eqnarray}
The projection of a solution $\gamma:I\to T^*M$ on $M$ is a geodesic with affine parametrization
in the special case when $H$ is the quadratic form corresponding to a Riemannian or pseudo-Riemannian metric.

In the following we assume that $H$
is invariant under the action of a connected Lie group $G$ on $T^*M$ induced by an action of $G$ on $M$.
Let  $\hat{Z}_a^* : T^*M\to TT^*M$, $a\in\mf{g}$, be the infinitesimal generator vector fields for the action of $G$ on $T^*M$. Their coordinate form is
\beq
\hat{Z}_a^*(x,p)=\frac{\partial\phi_a^i}{\partial\tau}(0,x)\frac{\partial}{\partial x^i}-
\frac{\partial^2\phi_a^j}{\partial\tau\partial x^i}(0,x)p_j\frac{\partial}{\partial p_i},
\eeq
where $\phi_a$ is the same object as in section \ref{sec.2}.\\

The invariance of $H$ implies the following {\it symmetry condition}: 
\beq
\label{SYMH}
\mathcal{L}_{\hat{Z}_b^*}H(x,p)=\frac{\partial H}{\partial x^i}(x,p)\frac{\partial \phi_b^i}{\partial\tau}(0,x)-
\frac{\partial H}{\partial p_i}(x,p)\frac{\partial^2\phi_b^j}{\partial\tau\partial x^i}(0,x)p_j=0,
\eeq
where $b\in\mathfrak{g}$. This equation holds for all $(x,p)\in T^*M$.\\

We recall that the {\it momentum map} for the action of $G$ on $T^*M$ is
$P: T^*M\to \mathfrak{g}^*,\ (x,p)\mapsto f_x^*(p)$, where $f_x$ is the linear mapping introduced in section \ref{sec.2} after theorem 2.3. Clearly $P$ is linear on 
each cotangent space $T_x^*M$, $x\in M$, and it is also equivariant.
$P$ restricted to the cotangent space $T_x^*M$ at $x\in M$ is the transpose 
of $f_x$.
$P$ has the  
property that
\beq
X_{(P|a)}=\hat{Z}_a^*\qquad \forall a\in\mathfrak{g},
\eeq
where $X_F$ denotes the Hamiltonian vector field corresponding to the function $F:T^*M\to \mathbb{R}$ and $(P|a)$ denotes the function 
$(x,p)\mapsto ( f_x^*(p) | a)$. 
This property implies $[X_{(P|a)},X_{(P|b)}]=X_{(P|{[a,b]})}$, where $[,]$ on the left hand side denotes 
the Lie bracket of vector fields.
The functions $(P|a)$, $a\in \mathfrak{g}$, are conserved quantities, i.e.\ the function $P$ (and thus $(P|a)$, for all $a\in \mathfrak{g}$) is constant along the solutions of the Hamiltonian equations.\\

\noindent
{\bf Definition 3.1}\ 
An element $a$ of $\mf{g}$ is called a  {\it relative equilibrium vector} at $(x,p)\in T^*M$ if
 the orbit of the corresponding one-parameter subgroup starting  at $(x,p)$ 
is a solution of the Hamiltonian equations.\\

Since the momentum map is constant along the solutions of the Hamiltonian equations, if $a\in \mf{g}$ is a relative equilibrium vector at $(x,p)\in T^*M$, then $a$ is an element of the stabilizer subgroup of $P(x,p)$ with respect to the coadjoint action of $G$.  \\

\noindent
{\bf Lemma 3.2}\ {\it
Let $H: T^*M\to \mathbb{R}$ be a Hamiltonian function that is invariant under the action of a connected Lie
group $G$.  $a\in\mathfrak{g}$ is a relative equilibrium vector at  $(x,p)\in T^*M$  if and only if 
\beq
X_H(x,p)=\hat{Z}_a^*(x,p),
\eeq
or, equivalently,
\beq
d(H-(P | a))(x,p)=0,
\eeq
where $P$ is the momentum mapping for the action of $G$  on $T^*M$.}\\

\noindent
The proof of this lemma  can be found in  \cite{Abr-Mars} (proposition 4.3.7.), 
for instance. 

The following generalization of the geodesic lemma can be stated for 
homogeneous spaces with invariant Hamiltonians.\\

\noindent
{\bf Lemma 3.3 (Geodesic lemma)}\ {\it
Let $M=G/K$ be a homogeneous space and $H:\ T^*M\to \mathbb{R}$ a $G$-invariant Hamiltonian function.
An element $a\in\mf{g}$ 
is a relative equilibrium vector at $(o,p)$, where $o$ denotes the origin, if and only if
\beq
\label{ham1}
dH_o(p)=f(a)
\eeq
and
\beq
\label{sta}
(\ f^*(p)\ |\ [a,b]\ )=0\qquad  \forall b \in \mathfrak{g}
\eeq 
hold, where $H_o$ is $H$ restricted to $T_o^*M$.
(\ref{sta}) is equivalent to the condition that the one-parameter subgroup generated by $a$ is contained by the stabilizer subgroup of 
$f^*(p)\in\mathfrak{g}^*$ with respect to the coadjoint action of $G$.}\\

\proof Assume first that $a$ is a relative equilibrium vector. (\ref{ham1}) is just the first  of the two Hamiltonian equations at the initial point and in coordinate form it reads as follows:
\beq
\label{d1}
\frac{\partial \phi^i_a}{\partial\tau}(x,0)=\frac{\partial H}{\partial p_i}(x,p).
\eeq
The second Hamiltonian equation at the initial point is
\beq
\label{d2}
\frac{\partial^2\phi^j_a}{\partial\tau\partial x^i}(0,x)p_j=\frac{\partial H}{\partial x^i}(x,p).
\eeq
Substituting the left hand sides of (\ref{d1}) and (\ref{d2}) for the right hand sides of (\ref{d1}) and (\ref{d2}) 
in (\ref{SYMH}) gives
\beq
\label{d3}
p_j\left[\frac{\partial \phi_a^i}{\partial\tau}(0,x)\frac{\partial^2\phi_b^j}{\partial\tau\partial x_i}(0,x)-
\frac{\partial \phi_b^i}{\partial\tau}(0,x)\frac{\partial^2\phi_a^j}{\partial\tau\partial x_i}(0,x)\right]=0\qquad
\forall b\in\mathfrak{g},
\eeq
which is just the coordinate form of
the equation
\beq
(\ p\ |\ [Z_a,Z_b](o)\ )=0\qquad \forall b\in \mathfrak{g}.
\eeq
This is  equivalent to (\ref{sta}), because $[Z_a,Z_b](o)=f([a,b])$ and 
$(p\ |\ f([a,b]))=(f^*(p)\ |\ [a,b])$. Considering the reverse direction,
it is clear that (\ref{d2}) can be obtained from  (\ref{d3}), 
(\ref{d1}) and (\ref{SYMH}).\ \ \ $\Box$\\

\noindent
{\bf Proposition 3.4 }\ 
{\it The set of relative equilibrium vectors at any point $(o,p)$ is an affine subspace of $\mf{g}$.}\\

\proof For any fixed $p$ the equations (\ref{ham1}) and (\ref{sta}) constitute an inhomogeneous linear system of equations for $a$, thus the solutions constitute an affine subspace in $\mf{g}$. $\Box$\\

A similar result holds for Lagrangian systems as well; in this case the statement is that the set of relative equilibrium vectors $a$ at $o$ for which $f(a)$ (which is the initial velocity of the orbit generated by $a$) is fixed is an affine subspace of $\mf{g}$.
This follows from the fact that the equations $f(a)=v$ and (\ref{c2}), where $v$ is fixed, constitute an inhomogeneous linear system for $a$.\\

\noindent
{\bf Definition 3.5}\ 
Let $M=G/K$ a homogeneous space and 
$H:T^*M\to\mathbb{R}$ a $G$-invariant Hamiltonian function. $(M,H)$ is a called a {\it Hamiltonian geodesic 
orbit (g.o.) space with respect to $G$}, if every solution of the Hamiltonian equations is an orbit of a one-parameter subgroup of $G$.\\

In the following propositions 3.6 and 3.7 elementary conditions are given under which a 
homogeneous space with an invariant Hamiltonian has the g.o. property.
They  
are direct consequences of lemma 3.2 and lemma 3.3.\\

\noindent
{\bf Proposition 3.6} {\it
Let $M=G/K$ be a homogeneous space and $H :\ T^*M\to \mathbb{R}$ a $G$-invariant Hamiltonian function. This dynamical system has the g.o.\ property with respect to $G$ if and only if
\beq
dH(o,p)\in \ \{ d(P|b)(o,p):b\in\mathfrak{g} \} \qquad \forall (o,p)\in T_o^*M
\eeq
or, equivalently,
\beq
X_H(o,p)\in \ \{ \hat{Z}_b^*(o,p):b\in\mathfrak{g} \} \qquad \forall (o,p)\in T_o^*M.
\eeq
}
\vspace{0.5cm}

\noindent
{\bf Proposition 3.7} {\it
Let $M$ and $H$ be the same as in the previous proposition. $(M,H)$ is a g.o.\ space with respect to $G$ if and only if 
for all $p\in T_o^*M$
there exists an $a\in\mathfrak{g}$
such that 
\beq
dH_o(p)=f(a)
\eeq
and 
\beq
(\ f^*(p)\ |\ [a,b]\ )=0\qquad  \forall b \in \mathfrak{g}
\eeq
hold.}\\

\noindent
{\bf Definition 3.8}\
Let $M=G/K$ be a Hamiltonian g.o.\ space with respect to $G$. A mapping $\xi: T_o^*M \to \mathfrak{g}$ with the property that $\xi(p)$ is a relative equilibrium vector at $(o,p)$ for all $p\in T_o^*M$ is called a {\it geodesic graph}. Obviously, there exists at least one geodesic graph for every Hamiltonian g.o.\ space.\\ 

The naturally reductive spaces are precisely those Riemannian g.o.\ spaces which admit a linear $K$-equivariant geodesic graph. If $M=G/K$ is naturally reductive and $\mf{g}=\mf{k}\oplus \mf{m}$ is a natural reductive decomposition, 
then the mapping $\xi$ defined as $\xi(p)=(f|_\mf{m})^{-1}(dH_o(p))$ is a linear 
$K$-equivariant geodesic graph. The mapping $p\mapsto dH_o(p)$ is a linear bijection between $T_o^*M$ and $T_oM$ in this case, since $H_o$ is quadratic and nondegenerate.
If $M$ is a Riemannian g.o.\ space and $\xi$ is a linear $K$-equivariant geodesic graph, then  
$\mf{g}=\mf{k}\oplus \xi(T_o^*M)$ is a natural reductive decomposition.
(See also the remarks  after definition 2.9.)\\

In the following last part of the section 
we describe the relation between g.o.\ spaces and $Ad^*(G)$-invariant functions, 
and we describe how an $Ad^*(G)$-invariant function that corresponds to a g.o.\ space 
can be used to obtain a geodesic graph. We present certain results on geodesic graphs and we discuss Riemannian g.o.\ spaces and naturally reductive spaces.  
We also describe a criterion that can be used to find  Hamiltonians or metrics that have the g.o.\ property. Finally, 
we discuss briefly a generalization of the notion of Hamiltonian g.o.\ space. \\
\newpage

\noindent
{\bf Lemma 3.9}\ 
{\it 
Let $M=G/K$ be a homogeneous space and $H :\ T^*M\to \mathbb{R}$ a $G$-invariant Hamiltonian function that has the g.o.\ property with respect to $G$.
If $P$ is constant along a smooth curve $\gamma: I\to T^*M$, then $H$ is also constant along this curve.}
\\

\proof The derivative $\frac{d (H\circ \gamma)}{d t}$ of $H$ along $\gamma$ at $t\in I$ equals  $(dH(\gamma(t))|\dot{\gamma}(t))$. It is sufficient to show that this number is zero for any $t\in I$. 
Let $t$ be a fixed element of $I$.
It follows from proposition 3.6. 
that $(dH(\gamma(t))|\dot{\gamma}(t))=(d(P|b)(\gamma(t))|\dot{\gamma}(t))$ for some $b\in \mathfrak{g}$.
Since $P$ is constant along $\gamma$, the derivative of $P$ along $\gamma$ is zero, therefore the derivative of $(P|b)$ is also zero, thus $(d(P|b)(\gamma(t))|\dot{\gamma}(t))=0$.
$\Box$
\\

The following theorem is a direct consequence of lemma 3.9.\\

\noindent
{\bf Theorem 3.10}\ 
{\it Let $M=G/K$ be a homogeneous space and $H :\ T^*M\to \mathbb{R}$ a $G$-invariant Hamiltonian function that has the g.o.\ property with respect to $G$.
If the connected components of the level sets of the momentum mapping $P$ 
have the property that any two point in them can be connected by a piecewise smooth curve, then 
$H$ is constant on the connected components of the level sets of $P$.
If, in addition, $H$ takes the same value on all connected components of any level set of $P$, then 
$H$ takes the form 
\beq
\label{hP}
H=h\circ P,
\eeq
where $h: \mathfrak{g}^* \to \mathbb{R}$ is an $Ad^*(G)$-invariant function.}\\
 
$P$ is an analytic function, therefore its rank is maximal on an open dense subset $N$ of $T^*M$, which is $G$-invariant. It follows that in $N$
the level sets of $P$  are submanifolds, therefore the condition of theorem 3.10 is satisfied and thus $H$ is constant on the connected components of the level sets of $P|_N$.

The formula $H=h\circ P$ always holds locally in $N$; if 
$(o,p)$ is in $N$, then there exists a suitable open neighborhood $O$ of $(o,p)$ in $N$ so that in this neighbourhood $H$ takes the form $H=h\circ P$, where $h$ is a (locally) $Ad^*(G)$-invariant smooth function on $P(O)$.
Furthermore, it follows from the proof of theorem 3.11, that
if $\dim P(O)=\dim G$, then  
$\xi: p'\mapsto dh(P(o,p'))$ is a smooth (locally) $K$-equivariant geodesic graph in an open neighbourhood of $p$ in 
$T_o^*M$. If $\dim P(O)<\dim G$, then $h$ can be extended to an open neighborhood of $P(O)$, and this extended version can be used to define $\xi$.
If the extension $\hat{h}$ of $h$ is $K$-invariant, then the local geodesic graph given by 
$\xi: p'\mapsto d\hat{h}(P(o,p'))$ is also $K$-equivariant.\\

The following theorem is a converse of theorem 3.10. Summation over the index $n$ is implied in the formulas (\ref{311a}) and (\ref{311b}).\\

\noindent
{\bf Theorem 3.11}\ 
{\it 
Let $h: \mathfrak{g}^* \to \mathbb{R}$ be an $Ad^*(G)$-invariant function with the properties that $h\circ f^*$ is smooth and $h$ is differentiable at the points of the image space of $f^*$ (which is $f^*(T_o^*M)$).
The Hamiltonian function defined as  
\beq
H=h\circ P
\eeq
is $G$-invariant and has the g.o.\ property. The vector 
\beq
\label{311a}
dh(P(o,p))=\frac{\partial h}{\partial g_n}(P(o,p))dg_n,
\eeq
where the $g_n$ are some linear coordinates on $\mathfrak{g}^*$, is a relative equilibrium vector at $(o,p)\in T^*M$, thus the mapping
\beq
\label{gm}
\xi=dh \circ f^*\, : T^*_oM \to \mathfrak{g},\ p\mapsto dh(P(o,p)) \equiv (dh\circ f^*) (p)
\eeq
is a $K$-equivariant geodesic graph.
}\\

\proof We note that $P(o,p)=f^*(p)$, by definition. $H$ is obviously $G$-invariant. The property that $h\circ f^*$ is smooth implies the smoothness of $H$. 
We have 
\beq
\label{311b}
dH=\frac{\partial h}{\partial g_n} \frac{\partial P_n}{\partial x^j} dx^j
+\frac{\partial h}{\partial g_n} \frac{\partial P_n}{\partial p_j} dp_j,
\eeq
where $P_n$ are the components of $P$ with respect to the coordinates $g_n$.
This shows that at $(o,p)\in T^*M$  the vector $b\in \mathfrak{g}$ that has the components $\frac{\partial h}{\partial g_n}(P(o,p))$ has the property that $dH(o,p)=d(P|b)(o,p)$, thus the condition of proposition 3.6 is fulfilled. Clearly $\frac{\partial h}{\partial g_n}(P(o,p))$ are just the components of $dh(P(o,p))$ with respect to the coordinates $g_n$.
$\Box$\\

It is also clear from the proof of theorem 3.11 that \\

\noindent
{\bf Proposition 3.12}\ 
{\it 
If $h: \mathfrak{g}^* \to \mathbb{R}$ is an $Ad^*(G)$-invariant function,
$H=h\circ P$ is a smooth Hamiltonian function and $h$ is differentiable at $P(o,p)$ for some $p\in T_o^*M$, then 
$dh(P(o,p))$ is a relative equilibrium vector at $(o,p)$.}\\

The condition imposed on $h$ in theorem 3.11 could probably be weakened, in particular we do not expect that the 
differentiability of $h$ in every point of $f^*(T_o^*M)$ is necessary for $h\circ P$ 
to be a g.o.\ Hamiltonian.

The following propositions 3.13-3.16, theorem 3.18, and partly theorem 3.17, are about Riemannian and pseudo-Riemannian spaces.\\

\noindent
{\bf Proposition 3.13}\ 
{\it 
If $h$ is a quadratic $Ad^*(G)$-invariant polynomial on $\mathfrak{g}^*$ and the polynomial $h\circ f^*$ is homogeneous, quadratic and nondegenerate, then $h$ gives rise to a Riemannian or pseudo-Riemannian g.o.\ metric on $M=G/K$.  
On $T^*_oM$ the quadratic polynomial that corresponds to the metric is $h\circ f^*$.
The geodesic graph $\xi: p\mapsto dh(P(o,p))$ is linear in this case.
If $h\circ f^*$ is positive definite, then the metric is naturally reductive.}\\

\proof
The Hamiltonian $H=h\circ P$ restricted to $T^*_oM$ is $h\circ f^*$, and the latter is a nondegenerate homogeneous quadratic polynomial, therefore $H$ corresponds to a Riemannian or pseudo-Riemannian metric. 
The quadraticity of $h$ implies that $dh$ is linear. $P(x,p)$ is also linear in the second variable, therefore $\xi: p\mapsto dh(P(o,p))$ is a linear map.  
If $h\circ f^*$ is positive definite, then the corresponding metric on $M$ is 
Riemannian. The linearity (and the $K$-equivariance) of $\xi$ implies, according to the remarks after definitions 2.9 and 3.8, that the metric is also naturally reductive. $\Box$\\

\noindent
{\bf Proposition 3.14}\ 
{\it 
If  $h$ is a smooth $Ad^*(G)$-invariant function on $\mathfrak{g}^*$ and $h\circ f^*$ is a homogeneous positive definite quadratic polynomial, then  $h$ defines a naturally reductive space.}\\

\proof 
$h$ gives rise to a Riemannian metric, since $h\circ f^*$ is a homogeneous positive definite quadratic polynomial.
$h$ is smooth, therefore we can take its quadratic part $h^{(2)}$ at $0\in \mathfrak{g}^*$.
$h^{(2)}$ is defined as $h^{(2)}(a)=\frac{1}{2}\sum_{n,m}\frac{\partial^2 h}{\partial g_n \partial g_m}(0)a_n a_m$,
where $g_n$ are linear coordinates on $\mf{g}^*$, $a\in \mf{g}^*$, and $a_n$ are the components of $a$ with respect to the coordinates $g_n$.
$h$ is $Ad^*(G)$-invariant and the action of $Ad^*(G)$ is linear, therefore 
$\frac{\partial^2 h}{\partial g_n \partial g_m}(0)$, as an element of $\mf{g}\otimes\mf{g}$, is a $G$-invariant tensor, and thus $h^{(2)}$ is also $Ad^*(G)$-invariant.
Moreover, $h\circ f^*=h^{(2)}\circ f^*$, since $f^*$ is linear and injective, thus $h^{(2)}$ gives rise to the same metric as $h$. As a consequence,  
$\xi: p\mapsto dh^{(2)}(P(o,p))$ is a $K$-equivariant linear geodesic graph, implying that the metric defined by $h$ is naturally reductive. $\Box$\\

The positive definiteness  of $h\circ f^*$ is not essential in the proof of this proposition; it is needed only to ensure that the metric to which $h$ gives rise is positive definite. The condition that $h$ is smooth can also be relaxed to the condition that $h$ is twice differentiable at $0$.\\

From a theorem of Kostant \cite{Kostant} generalized by D'Atri and Ziller \cite{DAtriZiller} it also follows
that all naturally reductive metrics can be 
obtained from $h$ functions that are  nondegenerate (not necessarily positive definite) quadratic polynomials. 
More specifically,  let $M=G/K$ be a naturally reductive Riemannian space with respect to $G$, 
$\mf{g}=\mf{k}\oplus\mf{m}$ a natural reductive decomposition, and assume that $G$ acts almost effectively on $M$ 
(i.e.\ the subgroup of elements that act as the identity transformation is discrete).
Then there exists an analytic subgroup $\bar{G}$ of $G$ and an analytic subgroup $\bar{K}$ of $K$ so that $M=\bar{G}/\bar{K}$ and the metric is naturally reductive with respect to $\bar{G}$ and it arises from an $Ad^*(\bar{G})$-invariant $h$ function that is a nondegenerate quadratic polynomial. 
The subgroup $\bar{G}$ is generated by $\bar{\mf{g}}=\mf{m}+[\mf{m},\mf{m}]$, which is an ideal in $\mf{g}$, and the Lie algebra of $\bar{K}$ is $\bar{\mf{k}}=\bar{\mf{g}}\cap \mf{k}$. 
The articles \cite{Kostant,DAtriZiller} also contain the result that if $h$ is a nondegenerate quadratic $Ad^*(G)$-invariant polynomial 
on $\mathfrak{g}^*$ and  $h\circ f^*$ is positive definite, then the metric defined by $h$ is naturally reductive.\\

\noindent
{\bf Proposition 3.15}\ 
{\it Let $M=G/K$ be a Riemannian or pseudo-Riemannian homogeneous space. If $a\in\mathfrak{g}$ is a relative equilibrium vector at $(o,p)$, then $\lambda a$ is also a relative equilibrium vector at $(o,\lambda p)$ for any $\lambda\in\RR$.}\\

\proof This result follows easily from lemma 3.3. $\Box$\\  
\newpage

\noindent
{\bf Proposition 3.16}\ 
{\it Let $M=G/K$ be a Riemannian or pseudo-Riemannian g.o.\ space.
If there exists a geodesic graph $\xi$ so that $\xi(0)=0$ and $\xi$ is differentiable at $0$, then there also exists a corresponding geodesic graph that is linear. If, in addition,  $\xi$ is $K$-equivariant, then the corresponding  
linear geodesic graph is also $K$-equivariant.}\\
 
\proof
The differentiability of $\xi$ and $\xi(0)=0$ imply that $\xi$ can be written as $\xi=\xi^{(1)}+\tilde{\xi}$, where $\xi^{(1)}$ is linear and $\tilde{\xi}$ 
has the property that $\lim_{\lambda\to 0} \tilde{\xi}(\lambda p)/\lambda =0$. $\xi^{(1)}$ is uniquely determined by $\xi$.  It follows from proposition 3.15 that 
$\xi_\lambda(p) =\xi(\lambda p)/\lambda $ is also a geodesic graph for any $\lambda > 0$. We have 
$\lim_{\lambda\to 0}\xi_\lambda(p)=\xi^{(1)}(p)$, thus $\xi^{(1)}$   is also a geodesic graph. If $\xi$ is $K$-equivariant, then obviously $\xi^{(1)}$ is also $K$-equivariant. $\Box$\\   

\noindent
{\bf Theorem 3.17}\ 
{\it Let $M=G/K$ be a Hamiltonian g.o.\ space. If $K$ is compact, then there exists a $K$-equivariant geodesic graph. If $K$ is compact and the space is Riemannian, then there exists a $K$-equivariant geodesic graph $\xi$ with the property that $\xi(\lambda p)= \lambda \xi(p)$ for all $\lambda\in\RR$ (i.e.\ $\xi$ is first order homogeneous).}\\ 

\proof
Due to the compactness of $K$ there exists a positive definite $Ad(K)$-invariant scalar product $Q$ on $\mf{g}$. For any $(o,p)\in T^*_oM$, consider the set of all relative equilibrium vectors at $(o,p)$, which is an affine subspace of $\mf{g}$ according to proposition 3.4. 
Let the value of the geodesic graph at $p$ be that unique element of this affine subspace which has the smallest norm with respect to $Q$. Since $Q$ is $Ad(K)$-invariant, the geodesic graph defined in this way is obviously $K$-equivariant.  Taking into consideration proposition 3.15, it is also obvious that this geodesic graph has the property $\xi(\lambda p)=\lambda \xi(p)$ for all $\lambda\in\RR$ 
if the Hamiltonian defines a Riemannian metric. $\Box$\\

It is easy to see that a similar theorem with a similar proof holds for Lagrangian g.o.\ spaces as well.
An $Ad(K)$-invariant scalar product on $\mf{g}$ exists also if 
$M=G/K$ is a Riemannian g.o.\ space and the action of $G$ on $M$ is effective (i.e.\ the compactness of $K$ is not necessary. See e.g.\ \cite{KSZ}, proposition 1 for a proof.) 
The proof of proposition 3.16 also shows that if a geodesic graph $\xi$ has the properties that $\xi(\lambda p)=\lambda\xi(p)$ for any $\lambda \in\RR $ and it is differentiable at $0$, then $\xi$ is linear. Consequently, we can state the following theorem:\\

\noindent
{\bf Theorem 3.18}\
{\it Let $M=G/K$ be a Riemannian g.o.\ space and assume that the action of $G$ on $M$ is effective. Then there exists at least one $K$-equivariant geodesic graph $\xi$ with the property that $\xi(\lambda p)=\lambda\xi(p)$ for any $\lambda \in\RR$. 
If $\xi$ is differentiable at $0$, then $\xi$ is linear and thus $M$ is a naturally reductive space with respect to $G$.
}\\

A geodesic graph that have the stated properties can be constructed in the same way as in the proof of 3.17.
The main result in Szenthe's paper \cite{Szenthe1}, which he obtained for affine g.o.\ manifolds with torsion-free affine connection
and for compact $K$, is similar to theorem 3.18.   
Our construction of the $K$-equivariant geodesic graph is simpler than that given in \cite{Szenthe1} (constructions similar to that in \cite{Szenthe1} can also be found in \cite{K-V,KowNikc}). 
For further results on the geodesic graphs of Riemannian g.o.\ spaces we refer the reader to \cite{KowNikc,Dusek}. \\

In section \ref{sec.4} we discuss an example where $h$ is 
a complicated function, nevertheless $h\circ f^*$ is a homogeneous quadratic polynomial and it is also positive definite, thus $h$ still gives rise to a Riemannian metric on $G/K$. This metric has the g.o.\ property, but the geodesic graph, which is unique in this example on an open dense set, is not linear and is not differentiable at $p=0$, 
and the metric is not naturally reductive with respect to $G$, in accordance with theorems 3.17 and 3.18. In addition to the 
nondifferentiability at $0$, the geodesic graph is also discontinuous along a one-dimensional subspace (from which $0$ is excluded). 

As the example shows, in the Riemannian case 
the function $h$ is not necessarily simple even though $H|_{T_oM}\equiv H_o=h\circ f^*$, and thus 
also $h|_{\mathfrak{m}^*}$, where $\mathfrak{m}^*$ is defined as $\mathfrak{m}^*=f^*(T_o^*M)$,
is a quadratic polynomial. However,  $h|_{\mathfrak{m}^*}$ is sufficient for determining $H_o$ (since $H_o= h|_{\mathfrak{m}^*} \circ f^*$), and thus $H$. Therefore in order to specify a Riemannian g.o.\ space it is sufficient to specify the polynomial $h|_{\mathfrak{m}^*}$, for which we introduce the notation  $h_o=h|_{\mathfrak{m}^*}$. 
The g.o.\ property implies that there is an open dense subset $N_o$ of $\mathfrak{m}^*$ such that
at any point $b\in N_o$ the derivative of $h_o$ has to be zero in any direction 
$ad_a^*(b)$, where $a\in \mathfrak{g}$ is such that $ad_a^*(b)\in \mathfrak{m}^*$. That is to say, at any point 
$b\in N_o$ the equation  
\beq
\label{co}
( d h_o (b)\ |\ ad_a^*(b) )=0
\eeq
has to hold for all $a\in \mathfrak{g}$ for which $ad_a^*(b)\in \mathfrak{m}^*$. This equation can be used in practice for finding suitable $h_o$ functions, i.e.\ for finding g.o.\ metrics or g.o.\ Hamiltonians, or to test whether a given metric or Hamiltonian function has the g.o.\ property. In terms of $H_o$, $h_o$ is given as $h_o=H_o\circ (f^*)^{-1}$, of course.

The $Ad^*(K)$-invariance of $h_o$ is necessary and sufficient for the $G$-invariance of the Hamiltonian function defined by $h_o$. If  $a\in \mathfrak{k}$ and $b\in N_o$, then  $ad_a^*(b)\in \mathfrak{m}^*$, thus ({\ref{co}}) 
has to be satisfied. However, if $h_o$ is $Ad^*(K)$-invariant, then ({\ref{co}}) obviously holds if
$a\in \mathfrak{k}$. The condition ({\ref{co}}) is therefore interesting mainly for those elements $a$ of $\mathfrak{g}$ which are not in $\mathfrak{k}$. \\  

The construction of g.o.\ Hamiltonian functions as $H=h\circ P$ can be generalized in the following way.\\

\noindent
{\bf Theorem 3.19}\ 
{\it 
Let $M$ be a manifold and $P$ a mapping $T^*M \to \mathfrak{g}^*$, where $\mathfrak{g}$ is a Lie algebra of a Lie group $G$, with the property $[X_{(P|a)},X_{(P|b)}]=X_{(P|[a,b])}$ for all $a,b\in \mathfrak{g}$. Let $h$ be a smooth $Ad^*(G)$-invariant function. The Hamiltonian function $H=h\circ P$  is $G$-invariant with respect to $G$ in the sense that $H$ is 
constant along the integral curves of 
$X_{(P|a)}$ for all $a\in \mathfrak{g}$. Any integral curve of $X_H$ coincides with an integral curve of $X_{(P|a)}$ for some $a\in \mathfrak{g}$. In particular, the integral curve of $X_H$ starting at the point $(x,p)\in T^*M$ coincides with the integral curve of $X_{(P|a)}$, where $a=dh(P(x,p))$, starting at $(x,p)$.}\\

In a more general form of the theorem the condition that $h$ should be smooth could be relaxed.
Certain notable dynamical systems, for example the system of two pointlike bodies which interact by the Newtonian gravitational 
force (the Kepler problem) and the harmonic oscillator, 
admit a formulation in this framework with noncommutative groups $G$. 
Completely integrable systems can also be formulated in the framework of theorem 3.19 with commutative 
symmetry groups.

\section{Example}
\label{sec.4}

In this section we discuss the example when $G=SU(3)$ and $K=SU(2)$ in order to give an illustration to the second part of section \ref{sec.3}. The $SU(3)$-invariant metrics on  $SU(3)/SU(2)$, which is diffeomorphic to the sphere $S^5$, constitute a two-parameter family. These metrics were described e.g.\ in \cite{Ziller}, where a complete description of the homogeneous metrics on the spheres was given.
In \cite{K-V} it was found that all the $SU(3)$-invariant metrics on $SU(3)/SU(2)$ have the g.o.\ property, but 
only a one-parameter subfamily is naturally reductive with respect to $SU(3)$.
Further results, in particular concerning the geodesic graph, were obtained in \cite{KowNikc}.
We note that these metrics belong to the type of g.o.\ metrics which are naturally reductive with respect to a suitable larger symmetry group \cite{KowNikc}. This larger group is $U(3)$ in the present case, and the stability subgroup of the origin is $U(2)$.

The Lie algebras of $SU(3)$ and $SU(2)$ are the following: \\

$su(3)=\mathfrak{g}=\mathfrak{k}\oplus\mathfrak{m}$

$su(2)=\mathfrak{k}=\mathrm{span} (A,B,C)$

$\mathfrak{m}=\mathrm{span} (E_1,E_2,E_3,E_4,Z)$

$$
\begin{array}{lllll}
[A,B]=2C & [A,Z]=0 & [A,E_1]=-E_2 & [B,E_1]=E_3  & [C,E_1]=E_4  \\ 
{}[B,C]=2A & [B,Z]=0 & [A,E_2]=E_1  & [B,E_2]=E_4  & [C,E_2]=-E_3 \\ 
{}[C,A]=2B & [C,Z]=0 & [A,E_3]=E_4  & [B,E_3]=-E_1 & [C,E_3]=E_2  \\ 
{}         &         & [A,E_4]=-E_3 & [B,E_4]=-E_2 & [C,E_4]=-E_1 \\
 & & & & \\
{}[Z,E_1]=E_2     & [E_1,E_2]=Z-\frac{1}{3}A & [E_2,E_4]=\frac{1}{3}B & &  \\
{}[Z,E_2]=-E_1  & [E_1,E_3]=\frac{1}{3}B   & [E_3,E_4]=Z+\frac{1}{3}A  & & \\
{}[Z,E_3]=E_4  & [E_1,E_4]=\frac{1}{3}C    & & &\\
{}[Z,E_4]=-E_3  & [E_2,E_3]=-\frac{1}{3}C.  & & &
\end{array}
$$

There exists one (up to multiplication by a constant) quadratic homogeneous invariant polynomial on $su(3)$:
\beq
Y_1=a'^2+b'^2+c'^2+e_1^2+e_2^2+e_3^2+e_4^2+z^2,
\eeq
where 
$a', b', c', e_1, e_2, e_3, e_4, z$ denote the coordinates corresponding to the basis vectors $A'=\frac{A}{\sqrt{3}}, B'=\frac{B}{\sqrt{3}}, C'=\frac{C}{\sqrt{3}}, E_1, E_2, E_3, E_4, Z$ of $su(3)$. $Y_1$ defines a positive definite $Ad$-invariant quadratic form on $su(3)$, allowing the identification of $su(3)$ and $su(3)^*$ and implying the equivalence of the coadjoint and adjoint actions of $SU(3)$.
The basis $A', B', C', E_1, E_2, E_3, E_4, Z$ is orthonormal with respect to the quadratic form defined by $Y_1$.
We use the same notation for the corresponding
orthonormal basis in $su(3)^*$. $Y_1$  can now be taken as an invariant polynomial on $su(3)^*$ 
as well.
$f$ can be used to identify $T_oM$ with $\mathfrak{m}$, and then
the momentum mapping restricted to $T^*_oM$, i.e.\ $f^*$, is the trivial 
embedding $\mathfrak{m}\to \mathfrak{m}\oplus \mathfrak {k}$. The polynomial $Y_1$ composed with $f^*$ 
thus takes the form 
\beq
\label{y1}
y_1=Y_1\circ f^*=e_1^2+e_2^2+e_3^2+e_4^2+z^2,
\eeq
where we have introduced the notation $y_1$ for $Y_1\circ f^*$.
The metric on $SU(3)/SU(2)$ corresponding to $y_1$ is naturally reductive. 
In \cite{K-V} it was found that
the complete family of Riemannian g.o.\ metrics on 
$SU(3)/SU(2)$ is given on $T^*_oM\equiv \mathfrak{m}$ by 
\beq
\label{metr}
\alpha (e_1^2+e_2^2+e_3^2+e_4^2)+\beta z^2,\qquad \alpha> 0,\ \beta> 0,
\eeq
where $\alpha$ and $\beta$ are real numbers. 
The metric (\ref{metr}) is naturally reductive if and only if $\alpha=\beta$ \cite{K-V}, which corresponds to $h=\alpha Y$. 
The family of polynomials (\ref{metr}) coincides with the complete family of positive definite $Ad^*(K)$-invariant quadratic homogeneous
polynomials on $\mathfrak{m}$. It is not difficult to verify that the metrics (\ref{metr}) also satisfy the condition (\ref{co}).

By solving the partial differential equations that express the $Ad^*(G)$-invariance of a function we find that 
the $Ad^*(G)$-invariant functions are of the form $G(Y_1,Y_2)$, where $G$ is an arbitrary function of two variables and $Y_2$ is the homogeneous third order polynomial
\beq
Y_2=\sqrt{3}\,\sigma_3 +z(\sigma_2-2\sigma_1)+\frac{2}{3}z^3,
\eeq
where $\sigma_1$, $\sigma_2$ and $\sigma_3$ are the following $Ad^*(K)$-invariant polynomials:
\begin{eqnarray}
\sigma_1 & = & a'^2+b'^2+c'^2\\
\sigma_2 & = & e_1^2+e_2^2+e_3^2+e_4^2\\
\sigma_3 & = & a'(e_1^2+e_2^2-e_3^2-e_4^2)+2b'(e_1e_4-e_2e_3)-2c'(e_1e_3+e_2e_4).
\end{eqnarray}
We have 
\beq
\label{y2}
y_2=Y_2\circ f^* = z(e_1^2+e_2^2+e_3^2+e_4^2)+\frac{2}{3}z^3,
\eeq
where the notation $y_2$ is introduced for $Y_2\circ f^*$.
In order to get the $G$ function for which $G(Y_1,Y_2)\circ f^*$ equals (\ref{metr}) one has to solve the equations (\ref{y1}) and (\ref{y2}) for $e_1^2+e_2^2+e_3^2+e_4^2$ and $z$. This involves the solution of a third order algebraic equation, 
therefore the result is a complicated formula that we do not write here.
This example shows that 
the function $h$ 
(which is $G(Y_1,Y_2)$ in the present case) can be complicated even though $h \circ f^*$ is a quadratic polynomial.

The geodesic graph can be calculated directly by solving the equations in lemma 3.3 or in lemma 2.4, 
as is done in \cite{KowNikc} (it is the equation (\ref{r}) that is actually used);
it is not necessary for this to know $h$. The result, which can be found written explicitly  below in equation (\ref{gg}) and in \cite{KowNikc}, has a relatively simple form.
The geodesic graph can also be calculated from the formula $\xi=dh\circ f^*$, where the necessary derivatives of $h$ can be determined from (\ref{co}). As a third approach, 
one can utilize the knowledge of the invariant polynomials $Y_1$ and $Y_2$ to calculate $dh\circ f^*$. Here we calculate the geodesic graph in this way, using
(\ref{y1}), (\ref{y2}) and (\ref{metr}). We have 
\beq
d(G(Y_1,Y_2))=\frac{\partial G}{\partial Y_1}dY_1+\frac{\partial G}{\partial Y_2}dY_2,
\eeq
thus we have to calculate the partial derivatives of $G$. (\ref{metr}), (\ref{y1}) and (\ref{y2}) can be written as 
\begin{eqnarray}
\label{ee1}
G(y_1,y_2) & = & \alpha r^2+ \beta z^2\\
\label{ee2}
y_1 & = & z^2+r^2 \\
\label{ee3}
y_2 & = & \frac{2}{3}z^3 + zr^2,
\end{eqnarray} 
where 
\beq
r^2=e_1^2+e_2^2+e_3^2+e_4^2.
\eeq
We have
\begin{eqnarray}
\label{gy1}
\frac{\partial G}{\partial y_1} & = & \frac{\partial G}{\partial r} \frac{\partial r}{\partial y_1}
+\frac{\partial G}{\partial z} \frac{\partial z}{\partial y_1}\\
\label{gy2}
\frac{\partial G}{\partial y_2} & = & \frac{\partial G}{\partial r} \frac{\partial r}{\partial y_2}
+\frac{\partial G}{\partial z} \frac{\partial z}{\partial y_2}.
\end{eqnarray}
For $\frac{\partial G}{\partial r}$ and $\frac{\partial G}{\partial z}$ we obtain
\beq
\label{xx1}
\frac{\partial G}{\partial r}=2\alpha r \qquad \frac{\partial G}{\partial z}=2\beta z
\eeq
from (\ref{ee1}). The partial derivatives $\frac{\partial r}{\partial y_1}$,  $\frac{\partial r}{\partial y_2}$, $\frac{\partial z}{\partial y_1}$ and $\frac{\partial z}{\partial y_2}$ can be calculated by taking partial derivatives of the equations (\ref{ee2}) and (\ref{ee3}) with respect to $y_1$ and $y_2$, and then solving the obtained four equations
for  $\frac{\partial r}{\partial y_1}$,  $\frac{\partial r}{\partial y_2}$, $\frac{\partial z}{\partial y_1}$ and $\frac{\partial z}{\partial y_2}$.
The result is
\begin{eqnarray}
\label{xx2}
\frac{\partial r}{\partial y_1}=\frac{z^2}{r^3}+\frac{1}{2r}\qquad \frac{\partial r}{\partial y_2}=\frac{z}{r^3}\\
\label{xx3}
\frac{\partial z}{\partial y_1}=-\frac{z}{r^2}\qquad \frac{\partial z}{\partial y_2}=-\frac{1}{r^2}.
\end{eqnarray}
Taking into consideration (\ref{gy1}) and  (\ref{gy2}) and using the results (\ref{xx1}), (\ref{xx2}) and (\ref{xx3})
we obtain for $\frac{\partial G}{\partial y_1}$ and 
$\frac{\partial G}{\partial y_2}$ that 
\begin{eqnarray}
\frac{\partial G}{\partial y_1} & = & \alpha +(\alpha-\beta)\frac{2z^2}{r^2}\\
\frac{\partial G}{\partial y_2} & = & -(\alpha-\beta)\frac{2z}{r^2}.
\end{eqnarray}
$dY_1$ and $dY_2$ are straightforward to calculate, and
the result for the geodesic graph is 
\begin{eqnarray}
[dG(Y_1,Y_2) \circ f^*](e_1E_1+e_1E_2+e_3E_3+e_4E_4+zZ) = \hspace{2cm} \nonumber\\
 2 \alpha (e_1E_1 +e_2E_2+e_3E_3+e_4E_4)+2\beta zZ \nonumber\\
 +(\beta-\alpha)\frac{2\sqrt{3}z}{r^2}[ (e_1^2+e_2^2-e_3^2-e_4^2)A' \nonumber\\
 +2(e_1e_4-e_2e_3)B'-2(e_1e_3+e_2e_4)C'],
\label{gg} 
\end{eqnarray} 
which agrees with the result obtained in \cite{KowNikc}, if we take into consideration the differences between the definitions 
in this paper and in \cite{KowNikc}. One difference that is worth noting is that in \cite{KowNikc} the geodesic graph is defined in such a way that only the $\mf{k}$-component is kept, i.e.\ the obvious $2 \alpha (e_1E_1 +e_2E_2+e_3E_3+e_4E_4)+2\beta zZ$ part is subtracted. 

(\ref{gg}) is well defined on an open dense subset of $T_o^*M$, but it does not have well-defined values at $r=0$ if $\alpha\ne \beta$.
It can be verified using (\ref{ham1}) and (\ref{sta}) that at $zZ$ (i.e.\ when $r=0$) all vectors 
$2\beta zZ+ aA'+bB'+cC'$, $a,b,c \in \mathbb{R}$, are relative equilibrium vectors.  
The limit of (\ref{gg}) in the points characterized by $r=0$ and $z\ne 0$ depends on the path (assumed to lie in the domain where $r\ne 0$) 
along which the limit is taken, therefore the geodesic graph is necessarily discontinuous in these points.

Several other examples of Riemannian g.o.\ spaces can be found in the literature (see e.g.\ \cite{K-V,KowNikc,CSG1}), which would also be interesting to discuss in a similar way.

\section*{Acknowledgments}

I would like to thank J\'anos Szenthe for proposing this subject and for useful discussions, and
L\'aszl\'o Feh\'er for his comments on the manuscript.
I also thank the referees for their constructive comments and for pointing out the references \cite{Lacomba,Ziller}.

\end{document}